\documentstyle[preprint,aps,floats]{revtex}
\def\be{\begin{equation}}
\def\ee{\end{equation}}
\def\bea{\begin{eqnarray}}
\def\eea{\end{eqnarray}}
\def\bml{\begin{mathletters}}
\def\blea{\begin{mathletters}\begin{eqnarray}}
\def\elea{\end{eqnarray}\end{mathletters}}
\def\p{\partial}

\input{epsf.tex}
\begin{document}
\draft
\title{Quantum tunneling of superconducting string currents.}

\author{Jose J.\ Blanco-Pillado$^{1,2,}$\footnote{Email address: {\tt
J.J.Blanco-Pillado@damtp.cam.ac.uk}}, Ken D.\ Olum$^{1,}$\footnote{Email address:
 {\tt kdo@alum.mit.edu}} and Alexander Vilenkin$^{1,}$\footnote{Email address:{\tt 
vilenkin@cosmos2.phy.tufts.edu}}}

\address{$^{1}$Institute of Cosmology, 
Department of Physics and Astronomy, 
Tufts University, 
Medford, Massachusetts 02155}
\address{$^{2}$
DAMTP, CMS, University of Cambridge\\
Wilberforce Road, Cambridge CB3 0WA, United Kingdom.}

\date{February 2002}

\maketitle

\begin{abstract}%
We investigate the decay of current on a superconducting cosmic string
through quantum tunneling.  We construct the instanton describing
tunneling in a simple bosonic string model, and estimate the decay rate.
The tunneling rate vanishes in the limit of a chiral current.  This
conclusion, which is supported by a symmetry argument, is expected to
apply in general.  It has important implications for the stability of
chiral vortons.  
\end{abstract}


\narrowtext

\section{Introduction}

Our universe may contain cosmic strings which are relics of a symmetry
breaking phase transition at early times.  (For a review, see
\cite{AlexBook}.) In many models these strings can be 
superconducting\cite{Witten85},
and the resulting current can have important effects on the string
dynamics.  In particular, if the current is large enough, the pressure
due to the charge carriers can oppose the tension of the string and
lead to a stable ``vorton'' state \cite{Davis88-2}.

The string current depends on the winding number density of the phase
of a condensate field; in a loop the total winding number will be an
integer, and thus classically cannot change as long as this phase is
well-defined.  When the current becomes large enough, there is the
possibility of a "quench", where the condensate is driven to 0, and
thus the phase can unwind and current can be lost.  However, even
below this threshold current, there is the possibility that current
can be lost through quantum tunneling effects.  It is those effects
which we consider here.

The simplest field theory model that exhibits bosonic superconductivity
is of the form,

\be\label{eqn:Lagrangian}
{\cal L} = \overline{D_\mu \sigma} D^\mu\sigma + \partial_\mu\bar
\phi~\partial^\mu\phi 
-{1\over 4} f_{\mu\nu}f^{\mu\nu}- V(\sigma,\phi),
\ee
where 
\bea
f_{\mu\nu} &=& \partial_{\mu} Z_{\nu} - \partial_{\nu} Z_{\mu} \, ,\nonumber\\ 
D_{\mu} \sigma &=& (\partial_{\mu} -ie Z_{\mu})\sigma \, .
\eea
This model has a local $U(1)_{local}$ symmetry associated with the gauge field
$Z^\mu$ and a global symmetry $U(1)_{global}$ of the phase
transformations of the field $\phi$.
In order to have a superconducting string model,  we should also
impose that the interaction potential between the
two complex scalar fields, $\phi$ and $\sigma$ is such that the broken
 $U(1)_{local}$ symmetry gives rise to a cosmic string configuration 
and the $U(1)_{global}$ becomes broken only inside the string. 
 A bosonic condensate will then
be formed on the string, with massless Goldstone bosons playing the
role of ``charge carriers''.  Note however that in this model the
condensate field $\phi$ is not coupled to any gauge field, so the
current is neutral.

Low energy excitations of the condensate along a straight string in the
$z$ direction will have the form
\be\label{eqn:ansatz}
\phi(x,y,z,t) = \phi(x,y) \, \exp[i \theta(t,z)] \, .
\ee
Using this ansatz, we can write an effective action for the field $\theta$
by inserting Eq. (\ref{eqn:ansatz}) in the four dimensional action to get,
\be
S = \Sigma \int dz dt [(\partial_t \theta)^2 - (\partial_z \theta)^2]
\label{S}
\ee
where 
\be 
\Sigma = \int dx dy |\phi|^2 \, .
\ee

We can now define the neutral current associated with the field living
on the string worldsheet by\footnote{Here and below, Latin indices
take values 0 and 1, Greek indices take values from 0 to 3, and the
signature of the metric is $g_{00} > 0$ both in four dimensions and on
the string worldsheet.}
\be
J_a =\partial_a \theta,
\ee
so the current is a measure of the winding number per unit
length of the phase of the bosonic condensate on the string.

If we imagine the straight string as part of a large loop, we can
write the total winding number
\be\label{eqn:N}
N = {1\over {2 \pi}} \oint \partial_{\sigma} \theta \, d\sigma,
\ee
which is classically conserved. Nevertheless, this is not the 
whole story. It was already noticed by Witten in his original 
paper\cite{Witten85} on superconducting strings that quantum effects
 would be able to decrease the current. He argued that since 
the field would have to unwind in order to 
decrease the value of N, the process would have to involve some kind
of localized region on the string worldsheet where the absolute value
 of the field $\phi$ would vanish.

Several authors have estimated the decay rate of the current following similar
arguments\cite{Zhang87,Haws88}. In this paper we shall study the quantum
tunneling of the current by analytically computing the worldsheet instanton
for this process and the corresponding decay rate in a loop with a 
uniform current.

\section{Worldsheet Instanton}

As a first approximation we will study the process on a straight
string and will neglect any backreaction
effect on the profile of the string. Therefore, we will only be concerned
with the evolution of the field $\phi$ in a fixed background. Also, we will
not consider situations where the unwinding of the field takes
place differently at different locations in the plane transverse to
the direction of the string.  This enables us to integrate over
transverse directions, and use only the effective Euclidean action for
the field $\theta$,  
\be
S_E = \Sigma \int dz \, d\tau [(\partial_\tau \theta)^2 +
(\partial_z \theta)^2] 
\label{SE}
\ee
The Euclidean field equation for $\theta$ is then
\be
\p_a\p_a\theta=0.
\label{eqtheta}
\ee
The limits of validity of this approximation will be indicated below.

Let us first assume that the current is spacelike,
\be
J_a J^a <0.
\ee
Then we can choose a frame of reference where the charge density
vanishes, $J_0=0$ (pure current).
We are interested in studying the quantum tunneling process from an initial 
state of homogeneous pure current of winding $N$,
\be
\theta= {{2 \pi N}\over L} z,
\label{thetai}
\ee
where $L$ is the length of
the loop, to a final state of winding $N-1$. The corresponding initial
current
is therefore, 
\bea
J_0 &=& 0 \, ,\nonumber\\
J_z &=& {{2 \pi N}\over{L}} \, .
\eea

Semiclassically, the tunneling is described by a Euclidean instanton
solution $\theta(\tau,z)$ which approaches the initial configuration
(\ref{thetai}) at $\tau\to\pm\infty$.  The field configuration 
immediately after tunneling is given by $\theta(0,z)$.
  Since the winding along the $z$-axis
$\tau=0$ should be one unit less than that at $\tau\to\pm\infty$, it
is clear that the field $\theta(\tau,z)$ should have a vortex below
and an anti-vortex above the $z$-axis (or vice versa).  The
corresponding solution of Eq. (\ref{eqtheta}) is
\be
\theta (\tau,z) =\arctan\left({z\over{\tau-a}}\right)- 
\arctan\left({z\over{\tau+a}}\right) +  {{2 \pi N}\over L} z,
\label{thetasol}
\ee
where $2a$ is the vortex separation.
This solution is well known from hydrodynamics \cite{Lamb}, where
$\p_a\theta$ corresponds to the velocity of an incompressible
fluid.  The solution describes a vortex-antivortex pair  
on the background of a fluid of constant velocity. 
The vortex separation $a$ is chosen so that the two vortices are in 
equilibrium due to the balance between their
mutual attraction and the oppositely directed Magnus
forces exerted by the background.  This gives \cite{Lamb} 
\be
a = {{L}\over{4\pi N}}.
\label{a}
\ee
The equilibrium is unstable, as it
should be, since the instanton solution should have a negative 
mode \cite{Callan77} .

For our purposes, instead of the variable $\theta$ it will be
convenient to introduce the dual field $\chi$ defined by
\be
\partial_a\chi =\epsilon_{ab}\partial_b\theta\,.
\label{chitheta}
\ee
Like $\theta$, $\chi$ obeys Laplace's equation,
$\partial_a\partial_a\chi = 0$ away from the vortex cores, but a vortex in
$\theta$ becomes a source for $\chi$.  This can be used to write the
equation for $\chi$ in the whole space,
\be
\p_a\p_a\chi=2\pi\rho,
\label{chieq}
\ee
where the ``charge density'' $\rho$ is given by
\be
\rho = {\tilde\delta}({\bf x-x_+})-{\tilde\delta}({\bf x-x_-}).
\ee
Here, ${\bf x}$ is a vector in the $(\tau,z)$-plane,
 ${\bf x_\pm}=(\pm a,0)$ are the locations of the vortex centers,
and ${\tilde\delta}({\bf x})$ is a delta-like function normalized so
that 
\be
\int{\tilde\delta}({\bf x})d^2x=1.
\ee
The width of the function ${\tilde\delta}({\bf x})$ is comparable to
that of the vortex core and its precise shape will not be important
for our considerations.
We thus have an analogy with
2-dimensional electrostatics.  The problem to be solved is that of two
opposite charges of magnitude $1$ in a constant external field
of magnitude

\be
{\bf E}=(2\pi N/L,0).
\label{E}
\ee  
The corresponding solution of (\ref{chieq})
outside the sources is
\be
\chi =\ln {|{\bf x-x_+}|\over{|{\bf x-x_-}|}} + {\bf E \cdot x}.
\label{chisol}
\ee

\begin{figure}
\begin{center}
\epsfxsize=3in
\epsfysize=3in
\begin{picture}(300,200)
\put(160,1){\Large{$z~\rightarrow$}}
\put(40,120){\Large{$\tau$}}
\put(40,140){\Large{$\uparrow$}}
\put(60,20){\leavevmode\epsfbox{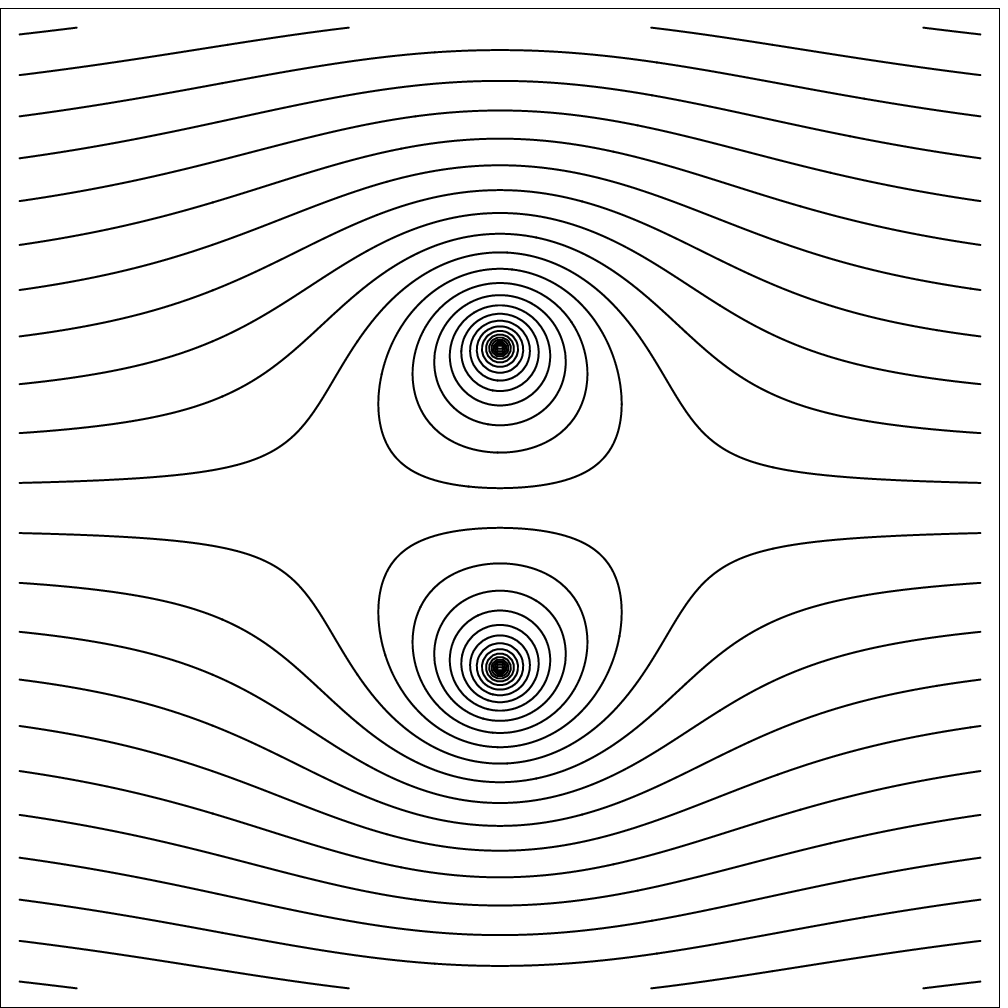}}
\end{picture}
\end{center}
\caption{ This figure shows the lines of constant
$\chi$ for the Euclidean instanton solution Eq. (\ref{thetasol}) on
the string worldsheet. The attractive force exerted between the vortex
and anti-vortex is balanced by the Magnus force due to the background current.}
\label{fig-1}
\end{figure}

The mutual attraction
of the charges must be balanced by the force of the external field acting
on them in opposite directions, which occurs when $E=1/2a$, in
agreement with (\ref{a}). See Fig \ref{fig-1}.

The Euclidean action corresponding to the field equation (\ref{chieq})
can be written as
\be
S_E =-\Sigma\int d^2x [(\p_a\chi)^2 +4\pi\rho\chi].
\label{S'}
\ee
The overall factor has been chosen so that the actions (\ref{SE}) and
(\ref{S'}) coincide when $\chi$ is a solution of (\ref{chieq}) and
$\theta$ is related to $\chi$ by Eq. (\ref{chitheta}). 

The current decay rate is given by
\be
\Gamma=A e^{-B},
\ee
where 
\be
B= S_E-S_E^{(0)},
\ee
$S_E$ is the instanton action and $S_E^{(0)}$ is the action for
the constant current configuration (\ref{thetai}).  
The semiclassical approximation
applies when $B\gg 1$.  In this regime, the tunneling 
rate is determined mainly by $B$, and the value of the prefactor $A$ is not very
important.  

\subsection{Evolution subsequent to the decay.}

\begin{figure}
\begin{center}
\epsfxsize=3in
\epsfysize=3in
\begin{picture}(300,200)
\put(220,70){\Large{$\rightarrow$}}
\put(100,90){\Large{$\leftarrow$}}
\put(-30,82){Charge density}
\put(280,82){$0$}
\put(280,200){$0$}
\put(-30,200){Winding density}
\put(60,20){\leavevmode\epsfbox{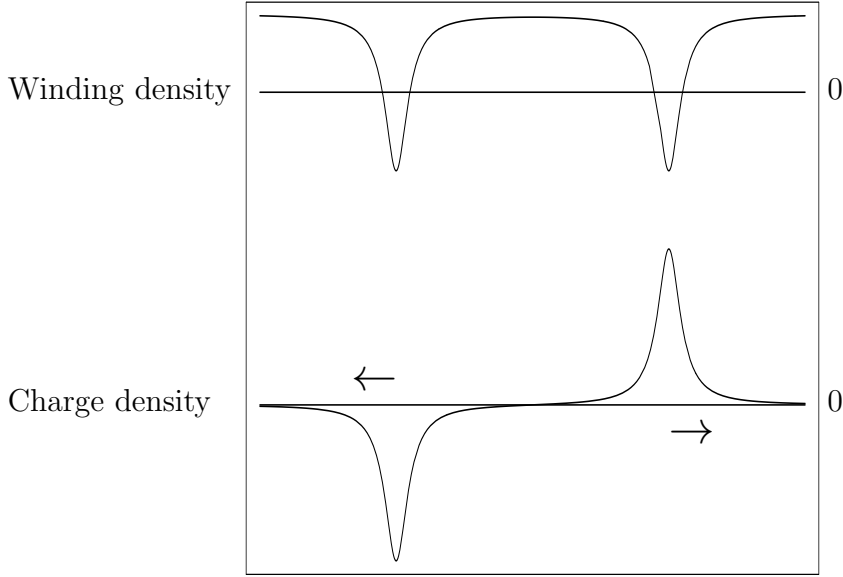}}
\end{picture}
\end{center}
\caption{Evolution of the winding and charge density in the string
condensate after the quantum current decay process. The two travelling
waves of opposite charge fly out from the center of the nucleation
process and their negative winding with respect to the positive
background decreases the overall integrated winding along the string. }
\label{fig-2}
\end{figure}

It is easy to check that our solution (\ref{thetasol}) conserves
energy and momentum as well as charge on the worldsheet. We can
also analytically continue it into Lorentzian time to obtain the
evolution of the current after the tunneling process has occurred.
By calculating the charge density on the Lorentzian worldsheet after 
the current decay, we can see (Fig. \ref{fig-2}) that this solution represents the 
propagation of two lightlike trains of opposite charge travelling in 
opposite directions along the string. They leave behind a region of
the same winding number density as the initial conditions, but if we
now calculate the total winding along the string we see that

\be\label{N1}
{1\over {2 \pi}} \oint \partial_{\sigma} \theta \, d\sigma = (N-1) \, .
\ee

This is because the two travelling waves have each negative
winding number density, in other words they have the opposite winding to the
original current.

\section{Decay rate}

In order to calculate the decay rate for this process we have to compute 
the Euclidean action for the instanton, or in terms of the
electrostatic analogy,
the difference in energy between the pure background configuration and 
the equilibrium solution described above.  However, both energies
are infinite, due to the infinite two-dimensional volume occupied by
the background electric field.  Moreover, one finds that a formal 
large-distance cutoff gives
different answers, depending on the details of the cutoff procedure.
We have therefore introduced a ``physical'' cutoff by replacing the
homogeneous electric field with a field produced by a finite
distribution of charges, $\rho_{ext}({\bf x})$.  The solution
(\ref{chisol}) is then replaced by
\be
\chi =\ln {|{\bf x-x_+}|\over{|{\bf x-x_-}|}} + \chi_{ext}({\bf x})
\equiv {\tilde\chi}({\bf x})+\chi_{ext}({\bf x}),
\label{chisol'}
\ee
where $\chi_{ext}({\bf x})$ is the solution of (\ref{chieq}) with the
source $\rho_{ext}({\bf x})$ and it is assumed that $\chi_{ext}({\bf
x})$ vanishes at $|{\bf x}|\to\infty$.  It is assumed also that the
external charges are located at distances $|{\bf x}|\gg a$, and in the
end we take the limit when these charges are removed to infinity,
while keeping the electric field near the origin fixed and equal to
(\ref{E}),
\be
-\int d^2 x\rho_{ext}({\bf x}){{\bf x}\over{r^2}}={\bf E}.
\label{E'}
\ee 

We also have to deal with the short-distance logarithmic divergence in
the self-energy of the two charges.  This divergence is an artifact of
our approximation treating the vortices as pointlike objects.  It is
easily fixed by replacing $\ln 0$ by $\ln \delta$, where $\delta$ is
the characteristic radius of the vortex core.  Although $\delta$ is defined
only by the order of magnitude, the result is not sensitive to the
precise definition, due to the logarithmic character of the divergence.

After integration by parts in Eq.~(\ref{S'}), the energy difference
between the instanton configuration and the background can be
expressed as
\be
B=-2\pi\Sigma\left(\int\rho\chi d^2 x 
-\int\rho_{ext}\chi_{ext} d^2 x\right) = 
-2\pi\Sigma[\chi({\bf x}_+)-\chi({\bf x}_-)] -2\pi\Sigma\int\rho_{ext}
({\bf x}){\tilde\chi}({\bf x}).
\label{S''}
\ee
Since the support of $\rho_{ext}({\bf x})$ is at $|{\bf x}|\gg a$, we
can use the large-distance expansion of ${\tilde\chi}$, 
\be
{\tilde\chi}\approx -{2{\bf x\cdot a}\over{r^2}},
\ee
in the last term of (\ref{S''}).  With the aid of Eq. (\ref{E'}), this
gives
\be
B= 4\pi\Sigma[\ln(2a/\delta)-2Ea],
\label{Sfinal}
\ee
where $\delta$ is the vortex core radius, as explained above.
Note that the result (\ref{Sfinal}) is insensitive to the details of
the external charge distribution $\rho_{ext}({\bf x})$. 

The action (\ref{Sfinal}) is maximized for $a_0=1/2E$, in agreement
with (\ref{a}).  It is easily verified that $d^2S/da^2(a_0)<0$, indicating
that $a=a_0$ is a local maximum of the action.  In terms of the
electrostatic analogy, this indicates that $a=a_0$ is an unstable
equilibrium, as it should be for an instanton.  Substituting $a=a_0$
in (\ref{Sfinal}) we obtain finally
\be
\label{bounce}
B= 4\pi\Sigma[\ln(L/2\pi N\delta)-1],
\ee
and for the tunneling rate
\be
\Gamma\sim A\left({2\pi N\delta\over{L}}\right)^K,
\label{rate}
\ee
where $K=4\pi\Sigma$.

The prefactor $A$ for tunneling in $(1+1)$ dimensions is given by
$A=BD$, where $D$ is a determinantal factor defined in Ref. \cite{Callan77}
and the factor $B$ comes from including the zero modes contribution to
the determinant. 
  
The exact calculation of this prefactor is a complicated task and we
do not attempt it here. Our main interest here is in the dependence of
the prefactor on the worlsheet current. In the Appendix we show that
we should expect this prefactor to scale as,
\be
A\sim B \,  \delta^{-2} \left({{L}\over{2\pi N \delta}}\right)^{2} \, .
\label{A}
\ee

Analytic estimates and numerical calculations show \cite{Hill87b} that,
if the fields $\sigma$ and $\phi$ are weakly coupled to each other and
to themselves, then $\Sigma\gg 1$ in a large portion of the parameter
space of the model (\ref{eqn:Lagrangian}).  In this case, $K\gg
10$, and it follows from Eqs. (\ref{rate}) and (\ref{A}) 
that the current decay is strongly suppressed, as long as the 
condensate winding does not get very large, namely,
$N\ll L/2\pi\delta$.

We finally discuss the validity of the approximations we made in
deriving the tunneling rate (\ref{rate}).  First, we  assumed that
the vortex cores give a negligible contribution to the instanton
action, compared to the  
contribution coming from the region 
\be
\delta\ll |{\bf x-x_{\pm}}|\lesssim a.  
\label{range}
\ee
This is justified if the logarithm
in Eq. (\ref{Sfinal}) is large, 
\be
\ln (L/2\pi N\delta) > 1, 
\label{largelog}
\ee
that is, when the current is very small.  We expect the vortex radius
$\delta$ to be comparable to the thickness of the condensate; then
Eq. (\ref{largelog}) says that
the wavelength of the condensate should be large compared to its
thickness.

We also assumed that the instanton solution can be factorized as in
(\ref{eqn:ansatz}), with $\theta$ independent of the coordinates
$(x,y)$ in the plane transverse to the string.  This approximation is
likely to break down in the vicinity of the vortex cores, where
$\theta$ varies on a length scale comparable to the string thickness, 
but we expect it to be valid in the range (\ref{range}) which,
assuming (\ref{largelog}), gives the dominant contribution to the
action.  Eq. (\ref{largelog}) is thus the only condition we need to
impose to justify the approximations made above.

\section{Charged current instanton.}

We would like now to extend the previous analysis to
the case where the condensate is coupled to the electromagnetic
field $A_{\mu}$. The simplest way to achieve this is to gauge the
$U(1)_{global}$ symmetry we have been discussing. In this case the
 action for the model takes the form,

\be
S_{el} =
\int{d^4x \, \overline{D_\mu \sigma} D^\mu\sigma 
-{1\over 4} f_{\mu\nu}f^{\mu\nu} +  \overline{{\cal D}_a \phi} {\cal D}^a\phi 
- V(\sigma,\phi) -{1\over 4}\, F_{\mu\nu} F^{\mu\nu}}
\ee
where 
\bea
F_{\mu\nu} &=& \partial_{\mu} A_{\nu} - \partial_{\nu} A_{\mu} \, ,\nonumber\\ 
{\cal D}_\mu \phi &=& (\partial_\mu -ie A_\mu)\phi \, .
\eea

As in the previous case, we are interested in the quantum mechanical
decay of a spacelike current of the form
\be 
\phi = \phi(x,y)\,  \exp[i \, \theta(\tau,z)] \,.
\ee

In terms of the angular variable the effective action becomes\cite{Witten85}
\be
S_{el} = \Sigma_{\text {eff}} \int dz d\tau (\partial_a \theta + e A_a)^2
 -{1\over 4}\int{d^4x \, F_{\mu\nu} F^{\mu\nu}} \, .
\ee
where $\Sigma_{\text{eff}}$ can be written in terms 
of the parameters in the original $U(1)_{local}\times U(1)_{local}$ 
 Lagrangian. This expression is correct if the scalar field condensate 
and the vector field remain approximately constant in the transverse
 direction to the string. This is ensured if the penetration depth of 
the vector field in the core of the string is large compared to the 
condensate thickness\cite{AlexBook}.

The important difference in this case is that the electromagnetic
 field is not confined to the worldsheet, so our 
instanton is somewhat more complicated. Adding the electromagnetic
 field makes the Euclidean equations of motion for the worldsheet 
look like those of a two dimensional superconducting sheet embedded 
in a four dimensional Euclidean space.
Our initial condition at $\tau = - \infty$ is a sheet carrying a
 constant planar current with its accompanying magnetic field. 
The instanton includes a vortex-antivortex pair in equilibrium under
 the action of the Magnus force, the electromagnetic force due to the
 long-range magnetic field of the vortices, and (to a smaller extent)
 the force due to the scalar field distribution on the sheet.

When this work was in progress we learned that this type of instanton
has been discussed by Duan\cite{Duan} in the context of a real
 superconducting wire. The instanton solution cannot be found
analytically in this case, but remarkably the instanton action,
(excluding the contribution from the vortex cores) can
still be found exactly\cite{Duan},

\be
B = {{4 \pi \Sigma}\over{\alpha}} 
\left[\ln \left[ 1 + \alpha \ln \, \left( {2a\over\delta} \right) \right] - 
 a \, \left({{4 \pi N \alpha}\over L}\right)\right]
\ee
where $\alpha = {{\Sigma e^2}\over {\pi}}$, and $N$, $L$, and $\delta$
have the same meaning as before. Note that this reduces to our
Eq. (\ref{Sfinal}) in the neutral limit $e \rightarrow 0$. The action is
maximized at
\be
a_0 \approx \left(L\over{4\pi N}\right) {{1}\over{ 1 + \alpha \ln \, \left( {L\over{2\pi N \delta}} \right)}}\,.
\ee
and finally it becomes
\be
B\approx {{4 \pi \Sigma}\over{\alpha}} 
\left[\ln \left[ 1 + \alpha \ln \, \left( {L\over{2\pi N \delta}} \right) \right] - 
{{\alpha}\over { 1 + \alpha \ln \, \left( {L\over{2\pi N \delta}}
\right)}}  \right]
\label{Bel}
\ee

As we discussed at the end of Section III, we are interested in the
range of parameters where $\Sigma > 100$ and $\ln (L/2 \pi N \delta) >
1$. With $e^2 \sim 10^{-2}$ this gives $\alpha > 1$ and Eq. (\ref{Bel})
reduces to

\be
B\approx {{4 \pi^2}\over{e^2}} 
\ln \left[\alpha \ln \, \left( {L\over{2\pi N \delta}} \right) \right]
\ee

Following the discussion for the neutral case we see that the
tunneling rate for the charged case is

\be
\Gamma\sim A \, \left( \alpha \,  \ln \left({L\over{2\pi N \delta}}\right)
\right)^{-(2\pi/e)^2}
\ee
which clearly indicates that this process is very much supressed in
the electromagnetic case as well.

\section{Discussion}

Our result (\ref{rate}) was derived in the frame where the global
charge density vanishes, $J_0=0$.  A generalization for an arbitrary
frame is easily obtained if we notice that the rate $\Gamma$ is Lorentz
invariant, so we only need to express it in terms of the invariant
quantity $J_aJ^a$.  With the aid of Eqs. (\ref{rate}), (\ref{A}) and
assuming (\ref{largelog}), we obtain
\be
\Gamma\sim K\delta^{-2}X^{(K-2)/2}\ln (1/X),
\label{Gammainv}
\ee
where
\be
X=|J_a J^a \delta^2|.
\ee

It is interesting to note that the rate ({\ref{Gammainv}) vanishes in
the chiral limit $J_a J^a=0$.  This is not surprising: since $\Gamma$
is an invariant, it can only depend on the invariant combination $J_a
J^a$ and should therefore vanish for a chiral current.  We expect this
conclusion to be valid for strings with charged as well as neutral
currents. The stability of chiral vortons with respect to this type of
decay reinforces the argument that cosmic string theories able to
produce chiral vortons will have serious constraints from their
overproduction in the course of the history of the string network\cite{Brandenberger,Carlos,Carter}.

For a timelike current, $J_a J^a>0$, instantons of the type we
discussed here do not exist.  In this case, one can go to the Lorentz
frame where the current vanishes and there is only a non-zero global
charge density.  Charge is conserved on the worldsheet and cannot
therefore be reduced by worldsheet instantons or any other processes
operating on the string worldsheet.  It is possible that when the
charge density gets high enough, charges can escape from the string by
tunneling into the external space.  We shall not consider such
tuneling processes in the present paper.

\section{Acknowledgments}

We would like to thank Jaume Garriga for very helpful
conversations. This work was supported in part by funding 
provided by the National Science Foundation. J.J.B-P was supported in
part by the Relativity Group PPARC Rolling Grant.

\section{Appendix}

In this Appendix we will estimate the dependence of the decay rate
prefactor on the uniform current on the worldsheet. The usual
procedure for this type of calculation involves the computation of 
the imaginary part of the energy coming from the 
contribution around the one-instanton solution of the Euclidean equations
 of motion. The decay rate per unit volume per unit time is given 
by\cite{Callan77},

\be
\Gamma = \prod_{i=1,2} \left({{N_i}\over{2\pi}}\right)^{1/2}
{\left|{Det'^{-1/2}{S''}_{E}}\right|\over{Det^{-1/2}{S''}_{E}^{(0)}}} \,
 e^{-(S_{E}-S_{E}^{(0)})}
\ee
where $S''_{E}$ and ${S''}_{E}^0$ denote the second order perturbation 
expansion of the instanton and the background Euclidean action respectively. 
In our notation we have

\be 
S_{E}-S_{E}^{(0)} = B \, .
\ee

Furthermore, the first term, proportional to $N_i$, is included to take into 
account the normalization of the zero modes, and in general it can be 
shown to be proportional to the instanton action $B$\cite{Callan77}.
Consequently we do not have to consider those in the 
following calculation. This is signaled by the expression
$Det'$ which denotes the determinant of the fluctuations once 
the zero modes have been excluded. In our system the zero modes correspond
to the collective motion of the vortex pair in the $(\tau,z)$ plane. It
is clear that translating them together does not alter the action, 
so this kind of fluctuations must correspond to zero modes.

Among the modes included in $Det'$ is the negative eigenvalue mode
of the fluctuations around the instanton solution. As we discussed
earlier, this negative mode corresponds to the fluctuation
that changes the distance between the vortices in the direction 
perpendicular to the current flow. 

There is also another type of fluctuation which corresponds to a 
 relative motion of the vortices in the direction
parallel to the current. This is a positive mode. 

If the current is very small, in other words, if the distance between the
two vortices is very large compared to the size of the vortex core, 
it could be expected that the only fluctuations which would not cancel
with the ones in the background are the fluctuations associated with
motions of the vortices. In this case the effective one-loop contribution
to the prefactor can be written as the following integral,

\be
\Huge{\Gamma} \sim B \, e^{-B} \int {{dz_r d\tau_r}\over{\mu^{-4}}} 
e^{-\beta({{z_r}\over{a_o}})^2} e^{-\beta({{\tau_r-a_o}\over{a_o}})^2} ,
\ee
where the factor of $B$ is the usual normalization factor for
the two zero modes in 1+1 dimensions. This expression singles out 
the contribution from the modes that are related to the position 
of the vortices by defining $z_r$ and $\tau_r$ as the relative distance
 between the vortices on the worlsheet coordinates. Furthermore, 
$a_0$ denotes the distance along the Euclidean time direction that 
maximizes the instanton action, $\beta=4\pi\Sigma$ and $\mu$ is a 
normalization factor of order
 $|\phi|\sim \delta^{-1}$ summarizing the result of doing 
the small fluctuation integral around the one vortex configuration. 

Performing the gaussian integrals we arrive at
\be
\Gamma \sim B \, e^{-B} \delta^{-2} \left({{L}\over{2\pi N \delta}}\right)^{2}
\ee


\begin{thebibliography}{1}

\bibitem{AlexBook}
A. Vilenkin and E. P. S. Shellard, {\em Cosmic Strings and other Topological
  Defects} (Cambridge University Press, Cambridge, 2000).

\bibitem{Witten85}
E. Witten, Nucl. Phys. {\bf B249},  557  (1985).


\bibitem{Davis88-2}
R. L. Davis and E. P. S. Shellard, Phys. Lett. {\bf B209},  485  (1988).

\bibitem{Zhang87}
S. Zhang, Phys. Rev. Lett. {\bf 59},  2111  (1987).

\bibitem{Haws88}
D. Haws, M. Hindmarsh, and N. Turok, Phys. Lett. {\bf B209},  255  (1988).

\bibitem{Lamb}
H. Lamb, {\em Hydrodynamics} (Cambridge University Press, Cambridge, 1932).

\bibitem{Callan77}
C. Callan and S. Coleman, Phys. Rev. {\bf D16},  1762  (1977).

\bibitem{Hill87b}
C. T. Hill, H. M. Hodges, and M. S. Turner, Phys. Rev. Lett. {\bf 59},  2493
  (1987).

\bibitem{Duan}
J.-M. Duan, Phys. Rev. Lett. {\bf 74},  5128  (1995).

\bibitem{Brandenberger}
R. Brandenberger, B. Carter, A. C. Davis and M. Trodden, Phys. Rev. 
{\bf D54}, 6059 (1996).

\bibitem{Carlos}
 C. J. A. P. Martins and E. P. S. Shellard, Phys. Lett {\bf B445}, 43 (1998).

\bibitem{Carter}
B. Carter and A. C. Davis, Phys. Rev. D {\bf 61}, 123501 (2000).


\end{thebibliography}

\end{document}